\documentclass{aa}  

\usepackage{natbib}
\bibpunct{(}{)}{;}{a}{}{,} 
\usepackage[breaklinks=true]{hyperref} 

\usepackage{graphicx}
\usepackage{txfonts}
\usepackage{xcolor}
\usepackage{enumitem}

\graphicspath{{figures/}} 
\def\msun{\!{\rm M}_\odot}
\def\lsun{\!{\rm L}_\odot}
\def\lsim{\mathrel{\rlap{\lower 3pt \hbox{$\sim$}} \raise 2.0pt \hbox{$<$}}}
\def\gsim{\mathrel{\rlap{\lower 3pt \hbox{$\sim$}} \raise 2.0pt \hbox{$>$}}}

\defcitealias{Venemans_et_al_2020}{V20}

\begin{document} 

\title{Are there more galaxies than we see around high-$z$ quasars?}


\author{Tommaso Zana\inst{1,2},
    Stefano Carniani\inst{1},
    David Prelogovi\'{c}\inst{1},
    Fabio Vito\inst{3},
    Viola Allevato\inst{1,3},
    Andrea Ferrara\inst{1},
    Simona Gallerani\inst{1},
    \and
    Eleonora Parlanti\inst{1}
    }

\institute{
Scuola Normale Superiore, Piazza dei Cavalieri 7, I-56126 Pisa, Italy\\
    \email{tommaso.zana@sns.it}
    \and
Dipartimento di Fisica, Sapienza, Universit\`a di Roma, Piazzale Aldo Moro 5, 00185, Roma, Italy
    \and
INAF - Osservatorio Astronomico di Bologna, via Piero Gobetti 93/3, I-40129 Bologna, Italy
    \and
INAF-Osservatorio astronomico di Capodimonte, Via Moiariello 16, I-30131 Napoli, Italy
    }
\date{Received June 13, 2023;}

\authorrunning{Zana T. et al.}

\abstract
   {Whether or not $z \gtrsim 6$ quasars lie in the most massive dark-matter halos of the Universe is still a subject of dispute.
   While most theoretical studies support this scenario, current observations yield discordant results when they probe the halo mass through the detection rate of quasar companion galaxies. 
   Feedback processes from supermassive black holes and dust obscuration have been blamed for this discrepancy, but the impact of these effects is complex and far from being clearly understood.}
   {This paper aims to improve the interpretation of current far-infrared observations by taking into account the cosmological volume probed by the Atacama Large Millimeter/submillimeter Array Telescope
   and to explain the observational discrepancies. 
   }
   {We statistically investigate the detection rate of quasar companions in current observations and verify if they match the expected distribution from various theoretical models, once convolved with the ALMA field-of-view, through the use of Monte Carlo simulations.}
   {We demonstrate that the telescope geometrical bias is fundamental and can alone explain the scatter in the number of detected satellite galaxies in different observations.
   We conclude that the resulting companion densities depend on the chosen galaxy distributions.
   According to our fiducial models, current data favour a density scenario where quasars lie in dark-matter halos of viral mass $M_{\rm vir} \gtrsim 10^{12}~{\rm M_{\odot}}$, in agreement with most theoretical studies. According to our analysis, each quasar has about 2 companion galaxies, with a [CII] luminosity $L_{\rm [CII]} \gtrsim 10^8~\lsun$, within a distance of about 1~Mpc from the quasar.}
   {}


\maketitle
   

\section{Introduction}

Most luminous ($L_{\rm bol} > 10^{46}$~erg~s$^{-1}$) quasars at redshift $z\gtrsim6$ are powered by the accretion process of gas onto the most massive supermassive black-holes ($10^{8}-10^{10}~\msun$; \citealt{Ferrarese_Ford_2005, Banados_et_al_2018, Wang_et_al_2018}).
According to theoretical models, including numerical simulations \citep{Sijacki_et_al_2009, Di_Matteo_et_al_2012, Costa_et_al_2014, Weinberger_et_al_2018, Barai_et_al_2018, Habouzit_et_al_2019, Ni_et_al_2020, Valentini_et_al_2021},
and clustering studies \citep{Hickox_et_al_2009, Ross_et_al_2009, Cappelluti_et_al_2010, Allevato_et_al_2011, Allevato_et_al_2012}, supermassive black-holes reside in the densest regions of the Universe, such as the centre of most massive dark-matter halos, with virial masses ranging from $10^{12}$ to $10^{13}~\msun$ at $z\sim6$.

However, observations do not always agree with such studies at nearly any redshift.
Numerous observational works have exploited current astronomical facilities to investigate the properties of the environments where quasars reside, yielding discordant results.
On the one hand, some studies suggest that quasars are located in regions characterized by large densities of galaxies, such as \citet{Steidel_et_al_2005} at $z=2.3$, \citet{Hall_et_al_2018} at $0.5 \lesssim z \lesssim 3.5$, \cite{Swinbank_et_al_2012} at $2.2 \lesssim z \lesssim 4.5$, \citet{Garcia-Vergara_et_al_2019} and \cite{Uchiyama_et_al_2020} at $z\sim4$, \citet{Husband_et_al_2013} and \citet{Capak_et_al_2011} at $z\sim5$, and \citet{Kim_et_al_2009}, \citet{Morselli_et_al_2014}, \citet{Balmaverde_et_al_2017}, \citet{Decarli_et_al_2017}, \citet{Decarli_et_al_2018}, \citet{Mignoli_et_al_2020}, and \citet{Venemans_et_al_2020} at $z \gtrsim 6$.
On the other hand, some works, such as \citet{Francis_Bland-Hawthorn_2004} and \citet{Simpson_et_al_2014a} at $z\gtrsim2$, \citet{Uchiyama_et_al_2018} at $z\sim4$, \citet{Kashikawa_et_al_2007} and \citet{Kikuta_et_al_2017} at $z\sim5$, and \citet{Banados_et_al_2013}, \citet{Mazzucchelli_et_al_2017b}, \citet{Champagne_et_al_2018}, \citet{Meyer_et_al_2022} at $z \gtrsim 6$, find that the number of companion galaxies is, at most, similar to the galaxy density estimated in blank fields.
Overall, the question of whether quasars tend to live in over-dense regions of the Universe remains an active area of research.

The detection of Lyman-break galaxies (LBGs) and Lyman-alpha emitting galaxies (LAEs) in the neighbourhoods of quasars and active galactic nuclei is often used in literature to probe the underlying dark matter distribution of high redshift structures (see, for instance, \citealt{Intema_et_al_2006, Venemans_et_al_2007, Banados_et_al_2013, Hennawi_et_al_2015, Mazzucchelli_et_al_2017b}, investigating the satellite number counts, or \citealt{Garcia-Vergara_et_al_2019}, studying the cross-correlation between galaxies and quasars).
The contradictory conclusions of these studies may also stem from the intrinsic challenges in constraining the redshift of LBGs and the significant influence of dust extinction and obscuration on both LBGs and LAEs.
For this reason, the best strategy to detect high-$z$ quasar companions relies on the observation of their [CII]158$\mu$m emission, as this gas tracer is not affected by dust-attenuation and intergalactic medium absorption \citep[e.g.,][]{Maiolino_et_al_2005, Walter_et_al_2009}.
In this context, the Atacama Large Millimeter Array (ALMA) has provided the most numerous spectroscopically confirmed
detections of $z\gtrsim6$ quasar companions \citep[][]{Decarli_et_al_2017, Decarli_et_al_2018, Venemans_et_al_2020}.
In particular, \citet[][hereafter \citetalias{Venemans_et_al_2020}]{Venemans_et_al_2020} 
found 27 line-emitter candidates within 27 quasar fields.
Out of this initial pool, 10 line emitters were already recognised as quasar companions in previous works \citep{Decarli_et_al_2017, Willott_et_al_2017, Neeleman_et_al_2019, Venemans_et_al_2019}. The remaining candidates have been identified as companions by assuming that the detected line does correspond to the carbon [CII] transition, emitted within $\pm \Delta v$ from the quasar.\footnote{In this work, we assume the validity of \citetalias{Venemans_et_al_2020} hypothesis and therefore consider all the additional 17 line emitters as proper quasar companions.}
Two thresholds have been adopted in \citetalias{Venemans_et_al_2020}, leading to slightly different results: 7 additional satellites are detected with $\Delta v = 1000$~km s$^{-1}$, and 9 with $\Delta v = 2000$~km s$^{-1}$. 
The \citetalias{Venemans_et_al_2020} sample is nowadays the largest and most complete catalogue of high-$z$ quasar companions with constrained redshift.

One debated possibility to explain the differences among these observational results is associated with the quenching effect of star formation in galactic satellites driven by quasar feedback \citep{Efstathiou_1992, Thoul_Weinberg_1996, Okamoto_et_al_2008, Dashyan_et_al_2019, Martin-Navarro_et_al_2019}.
According to this scenario, quasars would still reside in the most massive halos as predicted by theoretical models, but their companions would be gas-poor and would exhibit low star-formation activity, making them impossible to detect with current observational facilities.
This argument has been put forth to account for the absence of galaxy over-densities within $\sim1$~Mpc from certain high-z quasar objects.

\noindent
Nevertheless, this alleged negative feedback effect is currently far from established.
Some works proposed even a possible positive interaction between quasar feedback and star formation in companion galaxies \citep{Fragile_et_al_2017, Martin-Navarro_et_al_2021, Zana_et_al_2022, Ferrara_2023}.
In particular, \citet{Zana_et_al_2022}, based on cosmological simulations, and \citet{Ferrara_2023}, using semi-analytic models, claimed that quasar feedback always enhances the process of star-formation in satellites, provided that the satellites are not disrupted by the strong quasar outflows.
These recent works suggest that the discrepant observational results on the environments of high-$z$ quasars might not be due to feedback and that additional effects and possible observational biases might be in place.

\citet{Zana_et_al_2022} analysed the cosmological simulation by \citet{Barai_et_al_2018} and focussed on the environment of a massive quasar at $z\gtrsim6$, finding 2-3 satellites with a [CII] luminosity $L_{\rm [CII]}\gtrsim10^{8}~\lsun$.
They suggested that their findings match only the most densely populated field in \citetalias{Venemans_et_al_2020}, due to the geometrical effects introduced by the ALMA telescope. In fact, the relatively narrow ALMA field of view (FoV) compared with the much larger spatial range probed by the line of sight (LoS) where satellites can be identified, could inherently hinder the detection of a potentially significant fraction of the actual population of satellites, even if they are above the instrumental sensitivity threshold.
Other studies, such as \citet{Champagne_et_al_2018} and \citet{Meyer_et_al_2022} are in agreement with this hypothesis and conclude that observational campaigns adopting large FoV are fundamental to recover the missing satellite populations.

In this work, we investigate the geometrical bias introduced by ALMA on the satellite detection rate.
In particular, we use different mock distributions of galactic satellite populations, based on theoretically motivated models, to measure the geometrical response of the ALMA limited observational volume.
Our ultimate goal is to derive a rigorous statistical framework to interpret the most recent observational survey.

We assume a flat $\Lambda$CDM model, with the cosmological parameters $\Omega_{\rm M,0} = 0.3089$, $\Omega_{\Lambda,0} = 1-\Omega_{\rm M,0} = 0.6911$, and $H_{0} = 67.74$~Mpc s$^{-1}$ \citep[][results XIII]{Plank_2015}.
All lengths and volumes in the text are assumed to be physical unless otherwise specified.

The paper is structured as follows: we describe the method adopted to statistically investigate the geometrical bias in Sec.~\ref{sec:method}, in Sec.~\ref{sec:results} we present the outcomes, and in Sec.~\ref{sec:discussion} we discuss the results in the context of current literature.
We finally summarise our findings in Sec.~\ref{sec:conclusions}.

\section{Method}
\label{sec:method}

\begin{figure}
    \includegraphics[width=0.48\textwidth]{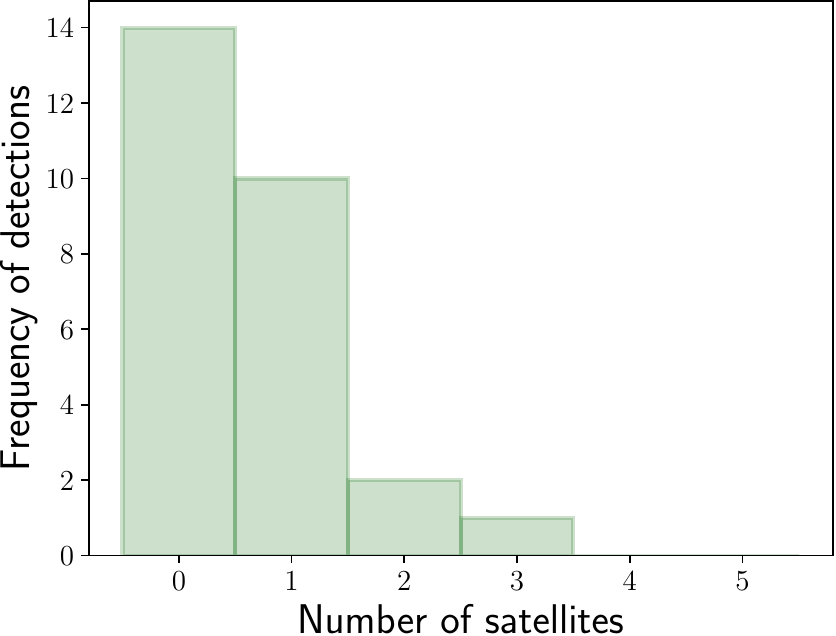}
    \centering
    \caption{Distribution of the number of satellites detected as [CII] emitters from \citetalias{Venemans_et_al_2020}, during an ALMA survey of 27 $z\gtrsim6$ quasars.
    All these sources reside within $\pm1000$~km~s$^{-1}$, i.e. about $\pm  1.3$~Mpc at $z=6.5$ from the central quasar.}
    \label{fig:hist_venemans}
\end{figure}

Figure~\ref{fig:hist_venemans} shows the incidence of [CII]-emitting companions in the sample of 27 quasars reported by \citetalias{Venemans_et_al_2020}, in the range $\Delta v = 1000$~km~s$^{-1}$.
Approximately half of quasars do not exhibit any detected companions, whereas in other cases, up to three galaxies are observed.
Since all datasets analysed by \citetalias{Venemans_et_al_2020} have comparable exposure times, the frequency of detections in this sample cannot be solely attributed to possible different sensitivities of the observations.
Under the assumption that all the observed quasars live in environments with similar properties and that the quasar feedback has the same effect on all the companions, our objective is to assess if the limited cosmic volume probed by ALMA can explain the large detection range shown in Figure~\ref{fig:hist_venemans}.

\begin{figure}
    \includegraphics[width=0.48\textwidth]{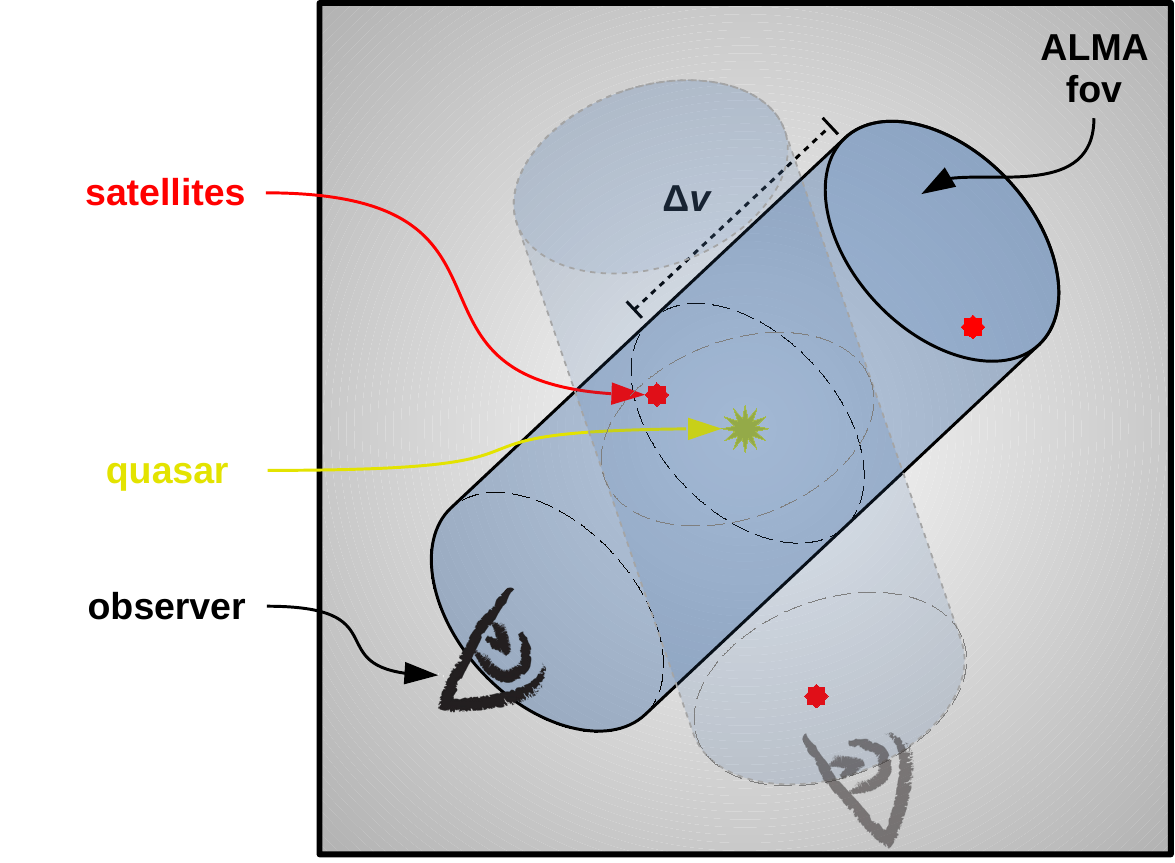}
    \centering
    \caption{Schematic representation of an ALMA observation of a quasar (yellow star). Here, two galaxy satellites (red stars) are detected, whereas a third one remains outside the observer's cylinder and would be detected only through a different line of sight.}
    \label{fig:scheme}
\end{figure}

ALMA observations of $z\sim6.5$ quasars are characterised by a circular FoV of radius $\simeq 90$~kpc and cover a wavelength range of $\Delta v = 1000-2000~{\rm km~s^{-1}}$ with respect to the systematic redshift of each quasar (\citetalias{Venemans_et_al_2020}).
The explored cosmological volume is thus a cylinder with a base of about $2.6 \times 10^{4}$~kpc$^{2}$ and a height of about 2.5-5.0~Mpc at $z=6.5$, depending on the spectral setting of the observations, as it is shown in Figure~\ref{fig:scheme}.
Here, we focus on the fixed LoS range $\Delta v = 1000$~km s$^{-1}$ to be more conservative, given the large distances involved, and we compare with \citetalias{Venemans_et_al_2020} equivalent sample.
Additionally, we consider the median redshift $z=6.5$ as representative of the range $6\leq z \leq 7$ for the whole \citetalias{Venemans_et_al_2020} sample.\footnote{The ratios amongst the comoving quantities (the position of the satellites and the ALMA observational range) are conserved in redshift. We have nonetheless run our analysis also at different mean redshift, confirming no significant variation.}

\noindent
Given the geometry of the probed volume, the detection rate of companions would depend on both the distribution of galaxies in the quasar environment and on the orientation of the cylinder axis.
Therefore, a fair comparison between theoretical studies and observations must take into account this effect.

We create different mock distributions of quasar satellites and study the frequency of detections when the companions lie within a random observational cylinder.
Ideally, we would run numerous, cosmologically motivated, N-body simulations of high-$z$ quasar companions and select a single observational cylinder in each simulation.
This approach would be prohibitive in terms of time and computational costs.
Alternatively, we run various sets of Monte Carlo simulations where we spawn, for each set, an increasing number of satellites $N$ in the quasar-centred sphere with radius $D_{z=6.5}$, where $D_{z=6.5}$ is the distance corresponding to the recession velocity $\Delta v = 1000$~km s$^{-1}$ at $z=6.5$, i.e. about 1.3~Mpc.
The $N$ satellites represent massive galaxies bright enough to be detected by ALMA as [CII] emitters, provided that they fall within the volume probed by the telescope.
They indeed represent only the “detectable” fraction of the whole satellite population and, in principle, could be connected to the total dark matter density where the quasar resides.
For each set of simulations (at fixed $N$), we intersect the quasar-centred sphere with the ALMA cylinder with a random orientation, as exemplified in Figure~\ref{fig:scheme}.
In particular, for each given $N$, we perform $10^5$ simulations, varying the position of the satellites, along with the orientation of the observational cylinder. 
Hence, we compute the probability $P(N_{\rm o}|N)$, that $N_{\rm o}$ satellites are included in the cylinder, and thus detectable by ALMA, as a function of $N$.
We adopt 9 different dark-matter distributions to generate the satellite populations:
\begin{itemize}
    \item \textbf{Homogeneous distribution}: satellites are generated by following a homogeneous distribution within the spherical volume, with fixed radius $D_{z=6.5}$.
    At $z=6.5$, the volume measures about $9$~Mpc$^3$.
    This model is dubbed as {\it Homogeneous} and represents the simplest case, with no assumption on the radial distribution.
    \item \textbf{Plummer model}: satellite galaxies are spawned by following a Plummer density profile \citep{Plummer_1911}, with cumulative distribution function (CDF)
    \begin{equation}
        P(r)=\frac{r^3}{\left(r^2 + a^2\right)^{3/2}},
    \end{equation}
    where $r$ is the radius from the central quasar and $a$ is the scale radius.
    We adopt two instances for this distribution by choosing $a=25$ and 100~kpc.
    These models are dubbed as {\it Plummer25} and {\it Plummer100}.
    \item \textbf{Hernquist model}: analogously to the Plummer model case, here satellites are generated by mimicking a Hernquist density profile \citep{Hernquist_1990}, through the CDF 
    \begin{equation}
        P(r) = \frac{r^2}{(a+r)^2}.
    \end{equation}
    This model, where we fix the scale radius $a=5$~kpc, is dubbed as {\it Hernquist5}.
    \item \textbf{NFW model}: we sample 2 Navarro-Frenk-White density profiles \citep{Navarro_et_al_1996}, dubbed as {\it NFW25} and {\it NFW100}, via the CDF
    \begin{equation}
        \begin{aligned}
            P(r) &= \mathcal{N}(a) \left[\ln\left(\frac{a+r}{a}\right)-\frac{r}{a+r}\right] \\
            \mathcal{N}(a) &= \left[\ln\left(\frac{a+r_{\rm cut}}{a}\right)-\frac{r_{\rm cut}}{a+r_{\rm cut}}\right]^{-1},
        \end{aligned}
    \end{equation}
    with scale radii $a$ equal to 25~kpc and 100~kpc, respectively.
    $\mathcal{N}(a)$ is a normalization constant with $r_{\rm cut}=D_{z=6.5}$.
    \item \textbf{CC function}: companion galaxies are randomly spawned by following the cross-correlation function quasar-galaxies $\xi(r)=\left(\frac{r}{a}\right)^{\gamma}$, with $a = 4.1$ comoving Mpc and $\gamma = 1.8$ from \citet{Garcia-Vergara_et_al_2019}.
    In particular, we adopt the CDF 
    \begin{equation}
        \begin{aligned}
            P(r)&=\mathcal{N} r^3 \left[\frac{1}{3} + \frac{1}{3-\gamma}\left(\frac{a}{r}\right)^{\gamma}\right] \\
            \mathcal{N} &= r_{\rm cut}^3 \left[\frac{1}{3} + \frac{1}{3-\gamma}\left(\frac{a}{r_{\rm cut}}\right)^{\gamma}\right]^{-1},
        \end{aligned}
    \end{equation}
    where $\mathcal{N}$ is a normalization constant.   
    This model -- dubbed as {\it CCF} -- wants to mimic only the shape of the galaxy distribution function around quasars. The density is not fixed and varies with $N$ in the Monte Carlo simulations.
    \item \textbf{Cosmological simulations}: finally, satellite galaxies are spawned by following the dark-matter distribution of cosmological zoom-in simulations of high-redshift quasars.
    In this work, we employ two different cosmological simulations, dubbed here with {\it CosmoSim1} \citep{Barai_et_al_2018} and {\it CosmoSim2} \citep{Valentini_et_al_2021}.
    In particular, satellite locations are randomly selected amongst the dark-matter potential wells in a $z=6.5$ snapshot of each simulation, after running \textsc{AMIGA} halo finder code \citep[][]{Knollmann_et_al_2009} with a minimum of 20 bound particles to define a halo.
    The dark-matter halos are considered only if their distance from the centre of the main halo is, at most, $D_{z=6.5}$. 
    {\it CosmoSim1} and {\it CosmoSim2} represent our fiducial models, for they have been derived from a complex framework specially designed to describe the environment of high-$z$ quasars.\footnote{We note also that only cosmological simulations, amongst our models, succeed in reproducing the complex filamentary structures expected to form during the collapse of primordial dark-matter fluctuations.}
    Although both {\it CosmoSim1} and {\it CosmoSim2} describe the evolution of two very massive dark-matter halos, their virial masses and radii are quite diverse, being 
    $M_{\rm vir} = 3.3\times10^{12}~\msun$, $R_{\rm vir} = 66$~kpc and $M_{\rm vir} = 1.2\times10^{12}~\msun$, $R_{\rm vir} = 47$~kpc at $z=6$, for {\it CosmoSim1} and {\it CosmoSim2}, respectively,\footnote{We define the virial mass as $M_{\rm vir}=M_{200}=\frac{4\pi}{3}200\rho_{c}R^{3}_{200}$, where $\rho_{c}$ is the critical density of the Universe and $R_{200}$ is the radius enclosing 200 times $\rho_{c}$.} and this increases the generality of the investigation.
    We also note that the feedback prescriptions for the active galactic nuclei are significantly different in the two simulations, and this has a non-negligible impact on the dark-matter distribution \citep[see the effect on the merger history in][]{Zana_et_al_2022}.
\end{itemize}

The models based on the Plummer, Hernquist, and NFW profiles adopt a set of parameters ($a$ and $\mathcal{N}$) to force their CDFs to have $P(r) \gtrsim 0.99$ for $r=D_{z=6.5}$. In the rare occasion where a galaxy is spawned at $r>D_{z=6.5}$, the extraction process is repeated.
Thanks to this expedient, the number density of companion galaxies within the spherical volume of radius $D_{z=6.5}$ is directly comparable amongst all the models.

We note that these models do not claim to completely cover all the possible satellite distributions, but rather aim to probe the implications of highly diverse environments.

\section{Results}
\label{sec:results}

\begin{figure*}
    \includegraphics[width=\textwidth]{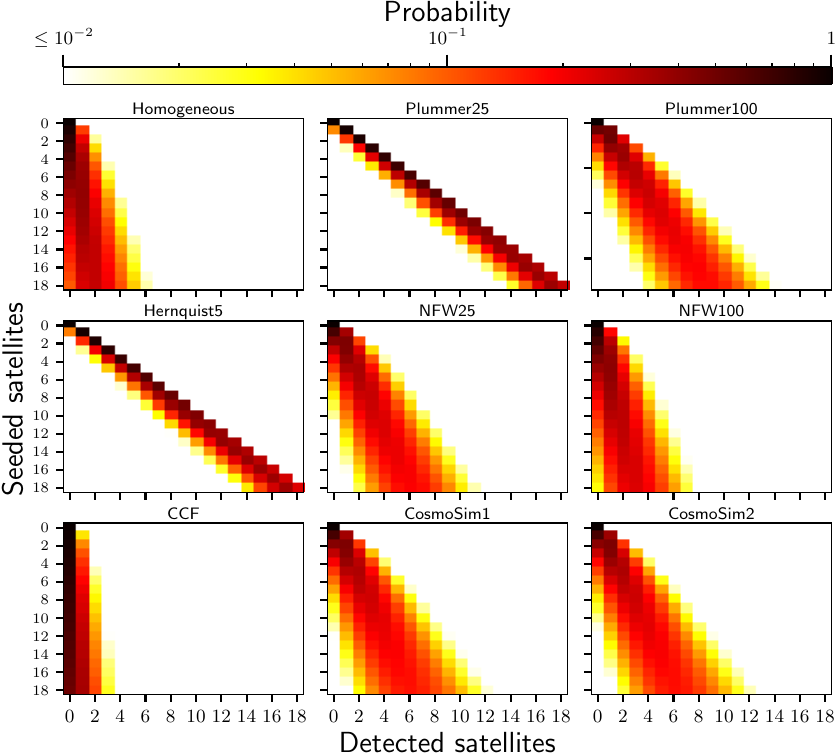}
    \centering
    \caption{Monte Carlo simulations for all the matter distribution models adopted, i.e. {\it Homogeneous}, {\it Plummer25}, {\it Plummer100}, {\it Hernquist5}, {\it NFW5}, {\it NFW15}, {\it CCF}, {\it CosmoSim1}, and {\it CosmoSim2}.
    Each panel shows the probability $P(N_{\rm o}|N)$ of detection of $N_{\rm o}$ satellites, given $N$ satellites spawned.}
    \label{fig:frequency}
\end{figure*}

Figure~\ref{fig:frequency} shows the outcome of the Monte Carlo simulations for the aforementioned nine satellite distributions.
In most cases, the number of detected companions is significantly smaller than $N$.
For example, in the fiducial distribution model {\it CosmoSim1}, with three seeded satellites, we estimate $P(N_{\rm o}=0|N=3)=0.29$. This indicates that the probability of no detection due to the ALMA FoV is $\sim30\%$, despite the presence of three satellites in the nearby environment.
We also observe that satellite distributions based on simulations {\it CosmoSim1} and {\it CosmoSim2} produce almost identical results, although they refer to two distinct quasar hosts.

\noindent
This hints at a major bias produced by the telescope observational volume that could explain why observations of quasar satellites return so different numbers in detecting serendipitous [CII] emitters.

The spherical geometry of {\it NFW25} produces a very similar trace to the simulation-oriented distributions and, therefore, could describe a similar distribution of galaxies, on average.
On the other hand, those distributions where satellites are almost always spawned in close proximity to the quasar, (i.e.{\it Plummer25}, or {\it Hernquist5}), result in a nearly one-to-one correspondence between seeded and detected satellites. 
Interestingly, if we distribute a variable number of companion galaxies by following an empirical cross-correlation function quasar-LAEs ({\it CCF}), galaxies are generated farther from the quasar and the detection efficiency drops at higher $N$. 
Finally, {\it Homogeneous} and {\it NFW100} yield similar outcomes to each other, even if they are in principle based on very different distribution laws.
It is worth mentioning that {\it Homogeneous} represents a control scenario, with no additional conjecture on the galaxy distribution.

\begin{figure*}
    \includegraphics[width=\textwidth]{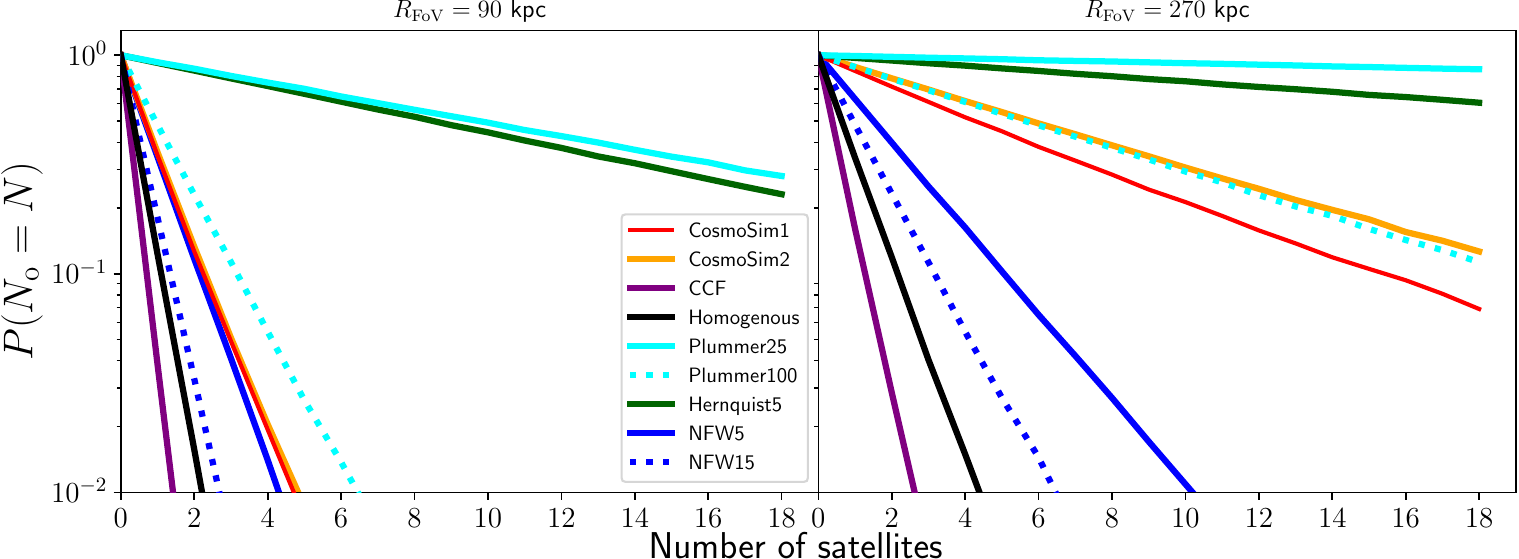}
    \centering
    \caption{Probability of total detection, $P(N_{\rm o}=N)$, for all the density profiles adopted in this work.
    {\it Left}: standard case of Figure~\ref{fig:frequency}, where $R_{\rm FoV}\simeq90~{\rm kpc}$.
    {\it Right}: mosaic-case with $R_{\rm FoV}\simeq270~{\rm kpc}$.}
    \label{fig:total_detection}
\end{figure*}

The left panel of Figure~\ref{fig:total_detection} shows $P(N_{\rm o}\!=\!N)$ for all the probed satellite populations.
As the seeded population grows, it becomes increasingly challenging to detect all the quasar satellites.
$P(N_{\rm o}\!=\!N)$ is always lower than 1 for $N>0$ and it decreases more rapidly the less compact the distribution is.
In Sec.~\ref{sec:discussion}, we discuss the possibility of sampling a cylinder with a radius three times larger using ALMA.
The right panel of Figure~\ref{fig:total_detection} presents the result for that case, indicating a much higher detection rate.

A general quasar environment is better quantified with a mean companion number, rather than their precise count, given the natural variance of a realistic system.
Therefore, we introduce a further step to take into account a scatter in the number of satellites that can be seeded.
In particular, we connect the number $N$ of seeded satellites to the average number $\langle N \rangle$ of a Poissonian distribution\footnote{The use of a Poissonian distribution in this context is a commonly made choice. We note that other scatter-laws could be adopted with very similar results.}
\begin{equation}
    P(N|\langle N \rangle)=\frac{\langle N \rangle^N}{N!}e^{-\langle N \rangle}.
    \label{eq:poisson}
\end{equation}
Hence, we can build the conditional probability
\begin{equation}
    P(N_{\rm o}, N | \langle N \rangle) = P(N_{\rm o}|N) \cdot P(N|\langle N \rangle),
    \label{eq:conditional_p}
\end{equation}
and the likelihood function for the average seeded number $\langle N \rangle$, given a single observation $N_{\rm o}$, $P(N_{\rm o}|\langle N \rangle)$: 
\begin{equation}
    P(N_{\rm o}|\langle N \rangle) = \sum_{N} P(N_{\rm o}, N | \langle N \rangle).
    \label{eq:conditional_p1}
\end{equation}

\begin{figure}
    \includegraphics[width=\columnwidth]{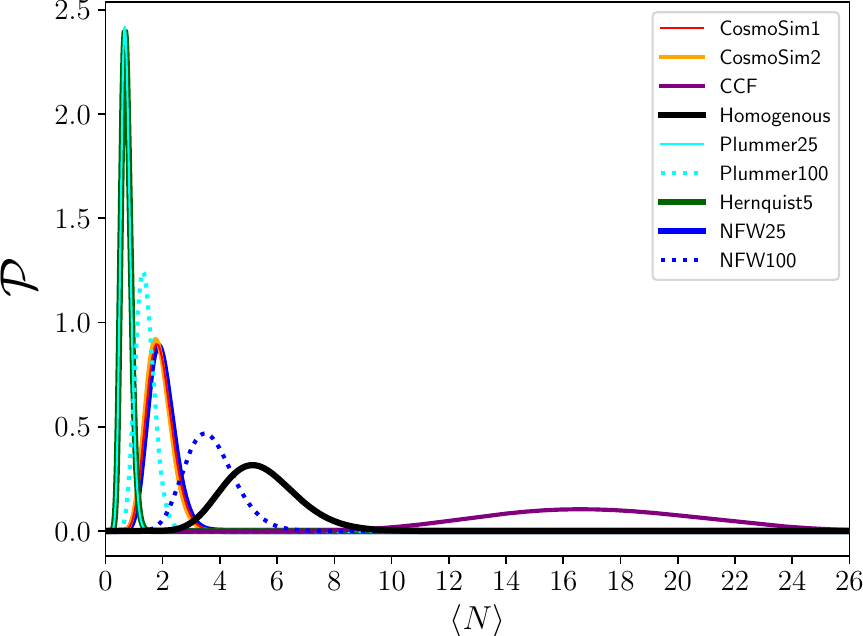}
    \caption{Posterior functions for the different galaxy distributions adopted in this work.
    The curves show the probability that an average number of massive satellites $\langle N \rangle$ is actually orbiting around a $z\gtrsim6$ quasar, given the observed data from \citetalias{Venemans_et_al_2020}.
    The colour coding is the same as Figure~\ref{fig:total_detection}
    }
    \label{fig:posterior}
\end{figure}

Finally, we compute the distribution $\mathcal{P}(\langle N \rangle)$ by including all the detections from \citetalias{Venemans_et_al_2020} and considering them to be equiprobable.
Operatively, we multiply together the likelihood associated with every detection and assume a flat prior over $\langle N \rangle$:
\begin{equation}
    \mathcal{P}(\langle N \rangle)=\frac{\prod_{N_{\rm o}} P(N_{\rm o}|\langle N \rangle)^{f(N_{\rm o})}}{\sum_{N_{\rm o}}\left[\prod_{N_{\rm o}} P(N_{\rm o}|\langle N \rangle)^{f(N_{\rm o})}\right]},
    \label{eq:likelihood}
\end{equation}
where $f(N_{\rm o})$ is the frequency of detection reported in Figure~\ref{fig:hist_venemans}.
The final likelihood functions, reproduced in Figure~\ref{fig:posterior} for all the distribution models, show that each model has a characteristic peak.
As mentioned before, the {\it CCF} model predicts the highest number of average intrinsic satellites in order to justify current observations.
On the other hand, compact distributions can describe observations with the smallest possible number of satellites.
In the specific cases of {\it Plummer20} and {\it Hernquist20} -- both sharply peaking around $\langle N \rangle=1$ -- \citetalias{Venemans_et_al_2020} observations with $N_{\rm o}=2$ or 3 would be explainable entirely through Poissonian fluctuations.

\section{Discussion}
\label{sec:discussion}

There is currently not enough observational data to constrain the spatial distribution of galactic companions around high redshift quasars.
For this reason, we have explored various possibilities, each with different degrees of realism and assumptions.
While numerical simulations describe the most realistic scenarios, the other radial profiles also offer a reasonable representation of reality, especially due to the limited number of objects involved.
Apart from the {\it CCF} model, which, nevertheless, has its roots in dedicated observational campaigns, all the spherically symmetric distributions adopted in this work are the most commonly used radial laws in astrophysics to model a gravitationally bound system, even if they are not directly connected to the scales and masses studied here.
\noindent
{\it Homogeneous} is the model with the fewest assumptions, where galaxies are randomly seeded in the surrounding volume.
The other spherically symmetrical models tend to refine the guess and distribute the galaxies preferentially closer to the central quasar.
{\it Plummer25}, {\it Plummer100}, {\it Hernquist5}, {\it NFW25}, and {\it NFW25} use different radial laws, with different scale radii.
Most compact distributions, such as {\it Plummer25}, {\it Plummer100}, and {\it Hernquist5} are less observationally motivated, especially at high $N$, where numerous companions are spawned within the virial radius of the quasar host, whereas the vast majority of \citetalias{Venemans_et_al_2020} satellites are detected at $r>100$~kpc.
On the other hand, the spherical model {\it CCF} describes a relation that is observed to hold for scales larger than $D_{z=6.5}$, beyond the limit of $1000$~km~s$^{-1}$.

More consistent models, such as {\it CosmoSim1}, {\it CosmoSim2}, {\it NFW25}, and {\it NFW100}, predict $\langle N \rangle \gtrsim 2$ on the basis of current observations.
This finding has numerous fundamental implications: the geometrical bias of ALMA can explain the scatter in recent observations.\footnote{This is valid even before the addition of the Poissonian scatter.}
In other words, we support the scenario in which high-$z$ quasars have an average of two massive companions, with [CII] luminosities $L_{\rm [CII]} \gtrsim 10^8~\lsun$, but their detection depends on the volume probed by the telescope and thus on the specific LoS of the observation.

\noindent
This result suggests that, for those quasar fields in which no satellite has been observed, mosaic observations with ALMA would result in the detection of 2 new galactic companions per quasar field, on average (in the cases described by our fiducial models, $\langle N \rangle \simeq 2$).
In the right panel of Figure~\ref{fig:total_detection}, we report the probability $P(N_{\rm o}\!=\!N)$, as a function of the number of seeded satellites if the FoV is expanded to $R_{\rm FoV}\simeq270~{\rm kpc}$, mimicking an observation with an additional layer of 9 adjacent ``standard'' FoV, around the original one.
For {\it CosmoSim1} and {\it CosmoSim2}, in particular, the probability of detecting the whole population of 2 seeded satellites is 0.72, which is almost six times larger than the probability estimated for a single ALMA pointing ($R_{\rm FoV}\simeq90~{\rm kpc}$).
If the intrinsic number of satellites $N$ were larger instead, $P(N_{\rm o}\!=\!N)\gtrsim 0.5$, up to $N \sim 6$.
Accordingly, follow-up priority should be given to those quasar fields which, so far, have shown the least amount of companions, given the high probability of detecting the whole missing population with an individual mosaic observation.

A further consideration can be drawn if we examine our fiducial model which predicts $\langle N \rangle \simeq2$, being in very close agreement with the twin model {\it CosmoSim2}.
In \citet{Zana_et_al_2022}, we found that a total of 2-3 satellites were detectable, having a [CII] luminosity higher than a current sensitivity threshold of about $10^{8}~\lsun$.
These results were based on a post-processing study of a suite of cosmological zoom-in simulations, including the evolution of baryons and numerous state-of-the-art sub-grid prescriptions.
In the present investigation, we have generalized the dark-matter distribution of satellites of such simulations (with no assumption on the mass of the halo) and demonstrated that current observations (\citetalias{Venemans_et_al_2020}) agree remarkably well with our previous prediction.
As a consequence, the dense quasar-host environment studied in \citet{Zana_et_al_2022}, and now confirmed by observations via our geometrical interpretation, likely describes real-quasar systems and this implies that most powerful high-$z$ quasars would live in the densest regions of the Universe.

Finally, we provide an additional interpretation by comparing our predicted number of satellites in the most realistic cases, with recent number-density measurements in a field galaxy population.
In the context of the large program ASPECS, \citet{Decarli_et_al_2020} and \citet{Uzgil_et_al_2021} have reported no detection of [CII]-emitters with $L_{\rm [CII]}>1.89 \times 10^8~\lsun$, in the redshift range $z\sim6-8$, within a comoving volume of about $12500$~Mpc$^3$, resulting in an upper limit on the galaxy density of $3.4 \times 10^{-4}$~Mpc$^{-3}$.
If we consider our predictions to take place within the quasar-centered sphere with radius $D_{z=6.5}$ (the volume enclosing an ALMA cylinder with every possible orientation), our fiducial models lead to a density $\gtrsim 0.2$~Mpc$^{-3}$, i.e., larger by almost a factor 600 with respect to the field environment probed by ASPECS.
Given that no [CII]-emitters have been found in the Hubble Ultra Deep Field via ASPECS, this result still holds true even if we compute our detected companions within a larger sphere, where the density could be reduced.

We conclude by observing that, the analysis conducted and its findings extend beyond ALMA, as they can be applied to any instrument with a limited FoV, due to the straightforward geometric nature of our study.

\section{Summary and conclusions}
\label{sec:conclusions}

Through various Monte Carlo simulations, we evaluated the geometrical bias of the ALMA telescope on the detection rate of high-$z$ quasar satellites (see Figure~\ref{fig:frequency}).
We convolved the resulting probability with a set of Poissonian distributions in order to estimate the intrinsic number of detectable companions, which are orbiting around a given quasar potential well.
We remark that each detectable satellite represents a galaxy massive enough to be detected as a [CII] emitter if it lies within the telescope geometry (see Figure~\ref{fig:scheme}).
We produced different likelihood functions for the average number of orbiting satellites, in order to explain the most recent ALMA observations (see Figure~\ref{fig:hist_venemans}).

We have demonstrated that:

\begin{enumerate}[label=(\it \roman*)]
\item The telescope bias can entirely explain the differences amongst the number of detected high-$z$ quasar companions.
Hence, every quasar could conceivably reside within almost the same environment, but we would detect only a part of the orbiting galaxies, depending on our line of sight.
\item If we consider the most physically motivated distribution profiles, e.g., {\it CosmoSim1} and {\it CosmoSim2}, we can infer an intrinsic number of $\sim2$ satellites, massive enough to be detected by ALMA as [CII] emitters with $L_{\rm [CII]} \gtrsim 10^8~\lsun$ and orbiting within $\sim 1.3$~Mpc from the quasar.
\item Interestingly, we expect to discover more satellites, in the case of {\it CosmoSim1} and {\it CosmoSim2}, via a simple mosaic observing campaign which targets those quasars where no companion galaxies have been observed so far.
We predict that a single additional layer of ALMA FoVs around those objects would be enough to observe the total population of $N\sim2$ satellites with a probability of 0.72, compared to 0.13 in the single-observation case.
In general, we expect such a campaign to detect, half of the time, the total population of companions, up to 6 satellites, i.e. $P(N_{\rm o}=N|N)\gtrsim 0.5$ for $N \lesssim 6$.
\item Our predicted number of satellites ($ii$) in the case {\it CosmoSim1} is compatible with the analysis performed in \citet{Zana_et_al_2022}.
In other words, ALMA observations are consistent with cosmological hydro-dynamical simulations evolving $z\sim6$ quasars in the most massive dark-matter halos with $M_{\rm vir} > 10^{12}-10^{13}~\msun$, corresponding to fluctuations of about $3-4\sigma$ in the density field.
\item We compared our fiducial predictions with a recent survey from the ASPECS program, where no [CII] emitters have been confirmed in the field above $L_{\rm [CII]}=1.89 \times 10^8~\lsun$ in our same redshift range, attesting once more the over-dense nature of high-$z$ quasar environments.

\end{enumerate}
Future additional ALMA detections of quasar companions will increase our current sample, allowing us to tune this model further and better constrain our predictions.
A contribution may also come from the AtLAST telescope \citep{Klaassen_et_al_2020} which, despite its low angular resolution, can detect high-$z$ satellites at a greater distance compared to ALMA.
Follow-up observations of \citetalias{Venemans_et_al_2020} galaxies and the possible detection of supplementary lines could confirm the status of companion or update the catalogues of quasar satellites.
We note, however, that the Poissonian scatter included in our analysis takes into account such potential small variations.
Moreover, the next deep JWST \citep{Gardner_et_al_2023} observational campaigns of quasar environments may also help to shed light on the spatial distribution of satellite galaxies, given the much larger FoV.
However, the galactic tracers probed will be completely different with respect to ALMA and will be subjected to dust extinction.
For this reason, a new and appropriate set of predictions is required.

\begin{acknowledgements}
    Funded by the European Union (ERC, WINGS, 101040227). Views and opinions expressed are however those of the authors only and do not necessarily reflect those of the European Union or the European Research Council Executive Agency. Neither the European Union nor the granting authority can be held responsible for them.
    TZ and VA acknowledge support from INAF-PRIN 1.05.01.85.08.
    The authors greatly thank the anonymous referee for useful comments which improved the quality of this manuscript.
\end{acknowledgements}

\bibliographystyle{aa} 
\bibliography{bibliography} 

\end{document}